\documentclass[10pt,fleqn]{article}
\usepackage{amssymb}

\newcommand{\name}[1]{\begin{flushleft}
                       \LARGE \bf #1
                       \end{flushleft}\vspace{-3mm}}

\newcommand{\Author}[1]{\begin{flushleft}
                       \it #1 \end{flushleft}}

\newcommand{\Adress}[1]{\begin{flushleft}
                       \it #1 \end{flushleft}}

\newcommand{\be}{\begin{equation}}
\newcommand{\ee}{\end{equation}}
\newcommand{\ba}{\hspace*{-5pt}\begin{array}}
\newcommand{\ea}{\end{array}}
\newcommand{\p}{\partial}
\newcommand{\ds}{\displaystyle}

\newcommand{\pbf}[1]{\mbox{\mathversion{bold}$#1$}}

\begin{document}

\name{On the three types of relativistic equations for particles with nonzero
mass}

\medskip

\noindent{published in {\it Lettere al Nuovo Cimento},  1972,  {\bf 4}, N~9, P. 344--346.}

\Author{Wilhelm I. FUSHCHYCH\par}

\Adress{Institute of Mathematics of the National Academy of
Sciences of Ukraine, \\ 3 Tereshchenkivska  Street, 01601 Kyiv-4,
UKRAINE}

\noindent {\tt URL:
http://www.imath.kiev.ua/\~{}appmath/wif.html\\ E-mail:
symmetry@imath.kiev.ua}

\bigskip

\noindent
In previous papers [1, 2] we have shown that there exist three types of the relativistic
equations for the massless particles. Here we show that for the free particles
and antiparticles with the mass $m>0$ and the arbitrary spin $s\geq \frac 12$
there also exist three types of nonequivalent equations.

For the sake of brevity we shall only dwell upon the equations of motion for the particles
with spin $s=\frac 12$. From the text it would be clear that all results of the paper can be
formulated for arbitrary spin. Let us consider the eight-component equation of the Dirac type~[3]
\be
\ba{l}
\ds (\Gamma_\mu p^\mu -\Gamma_4 m)\Psi(t,\pbf{x})=0, \qquad \mu=0,1,2,3,
\vspace{2mm}\\
\ds p_0 =i\frac{\p}{\p t}, \qquad p_a=-i\frac{\p}{\p x_a}, \quad a=1,2,3,
\ea
\ee
where the $8\times 8$ matrices $\Gamma_\mu$, $\Gamma_4$, $\Gamma_5$, $\Gamma_6$
obey the Clifford algebra; $\Psi$ is a eight-component wave function.

On the solutions of eq. (1) the generators of the Poincar\'e group $P_{1,3}$
have the form
\be
\ba{l}
\ds P_0\equiv {\mathcal H}=\Gamma_0\Gamma_a p_a +\Gamma_0 \Gamma_4 m\equiv
-2iS_{0k} p_k, \qquad p_4\equiv m, \quad k=1,2,3,4,
\vspace{2mm}\\
\ds P_a=p_a, \qquad J_{ab}=x_ap_b -x_b p_a +S_{ab},
\vspace{2mm}\\
\ds J_{0a}=x_0 p_a -\frac 12 (x_0 {\mathcal H}+{\mathcal H} x_a), \qquad
S_{\mu\nu}=\frac i4 (\Gamma_\mu \Gamma_\nu -\Gamma_\nu \Gamma_\mu).
\ea
\ee
Using the generators (2) it can be shown that on the set of solutions eq.~(1) a direct sum
of four irreducible representations of the group $P_{1,3}$:
\be
D^+_{s,0}\oplus D^-_{0,s} \oplus D^+_{0,s} \oplus D^-_{s,0}, \qquad s=\frac 12,
\ee
is realized. Here $D^\pm_{s,0}$ and $D^\pm_{0,s}$
denote the irreducible representations of the group $P_{1,3}$.
The symbols $D_{s,0}$ and $D_{0,s}$ denote the irreducible representations of the group
$O_4$. Elsewhere~[3] we have shown that eq.~(1) was invariant under the group
$O_6$, and a usual Dirac equation was invariant under the group $O_4$.

From (3) it follows that we can obtain three types of nonequivalent four-component equations
from eq.~(1). It is evident that these three types of equations are equivalent to one eq.~(1)
with three subsidiary conditions. These relativistic invariant subsidiary condition have
the form
\be
P_1^-\Psi =0 \quad \mbox{or} \quad P_1^+\Psi=0, \qquad P_1^\pm =\frac 12 (1\pm 2S_{56}),
\qquad S_{56}=\frac i2 \Gamma_5 \Gamma_6,
\ee
\be
P_2^-\Psi =0 \quad \mbox{or} \quad P_2^+\Psi=0, \qquad P_2^\pm =\frac 12 (1\pm 2\hat
\varepsilon S_{56}),
\qquad \hat \varepsilon =\frac{\mathcal H}{E},
\ee
\be
P_3^-\Psi =0 \quad \mbox{or} \quad P_3^+\Psi=0, \qquad P_3^\pm =\frac 12 (1\pm \hat
\varepsilon ), \qquad E=\sqrt{p_a^2+m^2}.
\ee
As the projective operators $P_a^\pm$ commute with the generators (2), it means that
subsidiary conditions (4), (6) are invariant under the Poincar\'e group.
The conditions (5), (6) are nonlocal in configuration space since
$\hat \varepsilon$ is the integrodifferential operator.

The eq. (1) together with the subsiduary condition (4) is equivalent to the usual Dirac equation.
In this case the wave function is transformed under the representation
\be
D^+_{s,0}\oplus D^-_{0,s} \quad \mbox{if} \quad P_1^-\Psi=0
\qquad \mbox{or} \qquad
D^-_{s,0}\oplus D^+_{0,s} \quad \mbox{if} \quad P_1^+\Psi=0.
\ee

Equation (1) together with (4) is equivalent to the four-component equation which coincide
on the form with the Dirac equation, however, the wave function in this equation is
transformed under the representation
\be
D^+_{s,0}\oplus D^-_{s,0} \quad \mbox{if} \quad P_2^-\Psi=0
\qquad \mbox{or} \qquad
D^-_{0,s}\oplus D^+_{0,s} \quad \mbox{if} \quad P_2^+\Psi=0.
\ee
It is clear that the representations (8) are not equivalent to (7).

Equation (1) with subsidiary condition (6) is equivalent to the four-component equation of the form
\be
i\frac{\p \Psi^{(4)}(t,\pbf{x})}{\p t} =E\Psi^{(4)} (t,\pbf{x}),
\ee
where the wave function $\Psi^{(4)}$ is transformed under the representation
\be
D^+_{s,0}\oplus D^+_{0,s} \quad \mbox{if} \quad P_3^-\Psi=0
\qquad \mbox{or} \qquad
D^-_{s,0}\oplus D^-_{0,s} \quad \mbox{if} \quad P_3^+\Psi=0.
\ee

It should be emphasized that only in the last equation of motion the Hamiltonian
is the positive operator. If we compare the particle with the representation $D^+_{s,0}$
and the antiparticle with the representation $D^+_{0,s}$,
then the eq.~(9) describes free motion of a particle and antiparticle with positive energy.
In this case the operator of a charge has the form $Q=\hat \varepsilon$.
Equation (1) with subsidiary conditions (4)--(6) can be written in the form
\be
(\Gamma_\mu p^\mu -\Gamma_4 m +\varkappa_a P_a^+)P_a^-\Psi(t,\pbf{x})=0,
\ee
where $\varkappa_a$ are the arbitrary constant numbers. For eq. (11) the conditions
(4)--(6) are automatically satisfied.

Equation (1) with the subsidiary conditions (4), (5), (6) (or three eqs. (11)) has different
$P$-, $T$-, $C$-properties. These properties can be read easily from the following
coupling scheme or irreducible representations of the Poincar\'e group
$$
\begin{array}{ccc}
D^+(s,0) & \stackrel{P}{\longleftrightarrow} & D^+(0,s)
\vspace{1mm}\\
\mbox{\footnotesize $T^p$} \updownarrow \mbox{\footnotesize $C$} & &
 \mbox{\footnotesize $C$} \updownarrow   \mbox{\footnotesize $T^p$}
\vspace{1mm}\\
 D^-(s,0) & \stackrel{P}{\longleftrightarrow} & D^-(0,s)\\
\end{array}
$$
$T^p$ is the Pauli $\leftrightarrow$ time-reversal operator. These questions will be considered
in more detail in another paper.

\medskip

\begin{enumerate}

\footnotesize

\item Fushchych W.I., {\it Nucl. Phys. B}, 1970, {\bf 21}, 321, {\tt quant-ph/0206077}; \\ {\it Theor. Math. Phys.}, 1971,
{\bf 9}, 91 (in Russian).

\item  Fushchych W.I., Grishchenko A.L., {\it Lett. Nuovo Cimento}, 1970,
{\bf 4}, 927, {\tt quant-ph/0206078};\\ Preprint ITF-70-88E, Kiev,
1970, {\tt quant-ph/0206079}.

\item Fushchych W.I., {\it Theor. Math. Phys.}, 1971, {\bf 7}, 3; Preprint ITF-70-32, Kiev, 1970.
 \end{enumerate}
\end{document}